\documentclass[preprintnumbers,prd,twocolumn,showpacs,amsmath,amssymb,superscriptaddress,nofootinbib]{revtex4}
\usepackage{bm}
\usepackage{graphicx}
\usepackage{hyperref}

\newcommand{\rmd}{\mathrm{d}}   
\newcommand{\rmi}{\mathrm{i}}   
\newcommand{\GF}{G_{\mathrm{F}}} 
\newcommand{\dmsol}{\Delta m_\odot^2} 
\newcommand{\dmatm}{\Delta m_\mathrm{atm}^2} 
\newcommand{\bfp}{\mathbf{p}}   
\newcommand{\bfq}{\mathbf{q}}   
\newcommand{\tot}{\mathrm{tot}}   
\newcommand{\vac}{\mathrm{vac}}   
\newcommand{\matt}{\mathrm{matt}}   
\newcommand{\Tr}{\mathrm{Tr}}   
\newcommand{\diag}{\mathrm{diag}}   
\newcommand{\sfH}{\mathsf{H}}   
\newcommand{\sfX}{\mathsf{X}}   
\newcommand{\bfH}{\mathbf{H}}   
\newcommand{\bmr}{\bm{\varrho}}   
\newcommand{\wpr}{\Omega_\mathrm{pr}}   
\newcommand{\kpr}{K_\mathrm{pr}}   
\newcommand{\hbe}{\hat{\mathbf{e}}}   
\newcommand{\nb}{n_\mathrm{b}}   
\newcommand{\tilf}{\tilde{f}}
\newcommand{\Ecsol}{E^\mathrm{s}_\odot}
\newcommand{\Ecatm}{E^\mathrm{s}_\mathrm{atm}}
\newcommand{\Ec}{E^\mathrm{s}_{2\times2}}
\newcommand{\wc}{\omega^\mathrm{s}}
\newcommand{\wcsol}{\omega^\mathrm{s}_\odot}
\newcommand{\wcatm}{\omega^\mathrm{s}_\mathrm{atm}}
\newcommand{\wtm}[1]{\tilde\omega_\mathrm{#1}}
\newcommand{\ntm}[1]{\tilde\nu_\mathrm{#1}}
\newcommand{\Hcr}{\sfH_\mathrm{cor}}

\begin{document}
\preprint{INT PUB 08-01}

\title{Stepwise Spectral Swapping with Three Neutrino Flavors}
\newcommand*{\INT}{Institute for Nuclear Theory, %
University of Washington, %
Seattle, WA 98195}
\newcommand*{\UCSD}{Department of Physics, %
University of California, San Diego, %
La Jolla, CA 92093-0319}
\newcommand*{\UMN}{School of Physics and Astronomy, %
University of Minnesota, Minneapolis, MN 55455}

\author{Huaiyu Duan}
\email{Huaiyu.Duan@mailaps.org}
\affiliation{\INT}
\author{George M.~Fuller}
\email{gfuller@ucsd.edu}
\affiliation{\UCSD}
\author{Yong-Zhong Qian}
\email{qian@physics.umn.edu}
\affiliation{\UMN}
\date{\today}

\begin{abstract}
We develop a framework 
for studying collective three-flavor neutrino oscillations
based on the density matrix formalism.
We show how techniques proven
useful for collective two-flavor  neutrino oscillations
such as corotating frames
can be applied  readily to  three-flavor mixing. 
Applying two simple assumptions and the conservation of two ``lepton numbers''
we use this framework to
demonstrate how the adiabatic/precession solution emerges.
We illustrate with a numerical example how
two stepwise spectral
swaps appear naturally if the flavor evolution
of the neutrino gas can be described by such a solution.  
For the special case where mu and tau flavor neutrinos
are equally mixed and are produced with identical energy
spectra and total numbers, we find that
one of the spectral swaps in the three-flavor
scenario agrees with that in the two-flavor scenario when appropriate
mixing parameters are used.
Using the corotating frame technique we show how the 
adiabatic/precession solution can obtain even in the presence of a
dominant ordinary matter background. With this solution we can explain
why neutrino
spectral swapping can be sensitive to  deviations from
maximal 23-mixing when the ``mu-tau'' matter term is significant.
\end{abstract}
 
\maketitle

\section{Introduction}

It has long been recognized that, in addition to the conventional
Mikheyev-Smirnov-Wolfenstein (MSW) effect
\cite{Wolfenstein:1977ue,Wolfenstein:1979ni,Mikheyev:1985aa},
neutrino self-coupling can be important for
neutrino flavor evolution when neutrino number densities are large
\cite{Fuller:1987aa,Notzold:1988kx,Pantaleone:1992xh,Sigl:1992fn,%
Fuller:1992aa,Qian:1993dg,Samuel:1993uw,Kostelecky:1994dt,Pastor:2001iu,%
Pastor:2002we,Balantekin:2004ug,Fuller:2005ae}.
Recently two-flavor neutrino oscillations in the core-collapse
supernova environment have been intensively investigated
\cite{Duan:2005cp,Duan:2006an,Duan:2006jv,Hannestad:2006nj,%
Raffelt:2007yz,EstebanPretel:2007ec,Duan:2007mv,Duan:2007bt,Fogli:2007bk}.
These studies show that supernova neutrinos can indeed
experience collective flavor evolution because of neutrino
self-coupling,
even when the neutrino self-coupling is subdominant compared
to the MSW potential \cite{Duan:2005cp}.

An important result is that collective two-flavor neutrino oscillations
can exhibit 
``stepwise spectral swaps'' or ``spectral splits'' 
in the final neutrino energy spectra when the neutrino number density
slowly decreases from a high value, where
neutrinos experience synchronized oscillations \cite{Pastor:2001iu}, 
towards zero
(see, e.g., Ref.~\cite{Duan:2006jv}). When this occurs,
$\nu_e$'s appear to swap their energy spectra with $\nu_x$'s
at energies below or above (depending on the neutrino mass hierarchy)
a transition energy $\Ec$ (where the superscript ``s'' can stand either
for ``swapping point'' or ``splitting point'').
Here $\nu_x$ is some linear combination of $\nu_\mu$ and $\nu_\tau$.
The phenomenon of spectral swapping is present
in both the ``single-angle approximation'', where flavor evolution
along various neutrino  trajectories is assumed to be the same as 
that along a representative trajectory (e.g., the radial trajectory),
and  the ``multi-angle approximation'', where flavor evolution
along different trajectories is independently
followed \cite{Duan:2006an}.
For the inverted neutrino mass hierarchy, it also has been found that
 stepwise neutrino
spectral swapping is essentially independent of the $2\times2$ effective
vacuum mixing angle when this angle is small 
(see, e.g., Ref.~\cite{Duan:2007bt}).

Collective two-flavor neutrino oscillations are best understood
with the help of neutrino flavor polarization vectors \cite{Sigl:1992fn}
or neutrino flavor isospins \cite{Duan:2005cp}. Using the spin analogy,
one can represent the flavor content of a neutrino mode by
a spin vector in  flavor isospace. In this analogy,
the matter effect is described as a spin-field coupling, and the
neutrino self-coupling is described as spin-spin coupling.
The phenomenon of spectral swapping is a result of collective
precession of all neutrino flavor isospins with a common angular velocity $\wpr$
at any given neutrino number density \cite{Duan:2006an,Duan:2007mv}.
This collective precession of neutrino flavor isospins
is described the two-flavor adiabatic/precession solution 
\cite{Raffelt:2007cb,Raffelt:2007xt}. In this solution, all neutrino
flavor isospins stay aligned or antialigned with a total effective
field in a reference frame that rotates with angular velocity $\wpr$.
Numerical simulations have shown that, in the supernova environment,
 neutrinos can first experience
collective MSW-like flavor transformation (in which the MSW effect
is enhanced by neutrino self-coupling) and then subsequently
the adiabatic/precession solution \cite{Duan:2007fw}.

In the real world, however, there are three active neutrino flavors.
Some limited progress has been made on understanding
collective three-flavor neutrino oscillations.
The first fully-coupled simulation of three-flavor neutrino oscillations in the
supernova environment showed a spectral swapping phenomenon similar
to the two-flavor scenario except possibly with two swaps \cite{Duan:2007sh}.
Another simplified numerical study with a single neutrino energy bin
\cite{EstebanPretel:2007yq}
reveals that collective neutrino oscillations can be sensitive
to deviations from maximal 23-mixing
when there is a dominant ``mu-tau'' term arising from
a high order contribution from virtual $\mu$'s and $\tau$'s
(e.g., Refs.~\cite{Fuller:1987aa,Botella:1987aa,Roulet:1995qb}).
Ref.~\cite{Dasgupta:2007ws} has extended the neutrino flavor polarization
vector notation to the three-flavor scenario and discussed
the collective three-flavor oscillations as  ``factorization''
of two two-flavor oscillations.

In this paper we develop a framework for studying collective
three-flavor neutrino oscillations based on the density matrix
formalism. Using this framework we find a generalized three-flavor
adiabatic/precession solution and show how stepwise
spectral swapping can appear as a natural result of such a solution.

The rest of this paper is organized as follows. In Sec.~\ref{sec:eom}
we develop a framework centered around 
a $3\times3$ reduced neutrino flavor density matrix. 
This density matrix is equivalent to  an 8-component neutrino flavor vector,
a generalization of neutrino flavor isospin and similar to
the Bloch vector used in Ref.~\cite{Dasgupta:2007ws}. We show
how techniques important
in studying collective two-flavor oscillations such as
corotating frames
can be applied in the framework.
In Sec.~\ref{sec:sol} we  demonstrate how the three-flavor
adiabatic/precession solution can be found using two simple assumptions
and the conservation of two ``lepton numbers''. We  also illustrate
with a numerical example how two spectral swaps can form
from the adiabatic/precession solution when the total neutrino number
density vanishes. In Sec.~\ref{sec:matt} we  employ the
corotating frame technique and show that the adiabatic/precession
solution obtains even in the presence of a dominant matter background.
In particular, we show that, in the presence of a large mu-tau
term, neutrino spectral swapping becomes sensitive 
to deviations from maximal 23-mixing. In Sec.~\ref{sec:conclusions}
we  give our conclusions.

\section{Equations of Motion%
\label{sec:eom}}

\subsection{Density Matrix Description}

The flavor content of a neutrino (antineutrino) mode with momentum $\bfp$
is generally described by  density matrix
$\rho_\bfp$  ($\bar\rho_\bfp$) \cite{Sigl:1992fn}.
The diagonal elements of a density matrix are the occupation numbers 
of the neutrino eigenstates in a particular basis, and the off-diagonal
elements contain the phase information relating to
 neutrino mixing. For a neutrino pure
state described by flavor wavefunction
\begin{equation}
\psi_{\nu_\bfp}=\begin{pmatrix}
a_{\nu_1}\\ a_{\nu_2}\\ a_{\nu_3}
\end{pmatrix},
\end{equation}
the density matrix is
\begin{equation}
\rho_\bfp=n_{\nu_\bfp}\begin{pmatrix}
|a_{\nu_1}|^2 & a_{\nu_1}a_{\nu_2}^* &  a_{\nu_1}a_{\nu_3}^* \\
a_{\nu_2}a_{\nu_1}^* & |a_{\nu_2}|^2 & a_{\nu_2}a_{\nu_3}^* \\
a_{\nu_3}a_{\nu_1}^* & a_{\nu_3}a_{\nu_2}^* &  |a_{\nu_3}|^2
\end{pmatrix},
\end{equation}
where $n_{\nu_\bfp}$ is the neutrino number density in momentum mode $\bfp$, and
$a_{\nu_1(\nu_2,\nu_3)}$ are the amplitudes for the neutrino
to be in the corresponding vacuum mass eigenstates. 
With this notation we have normalization
\begin{equation}
\sum_{i=1,2,3}|a_{\nu_i}|^2=1.
\end{equation}
The density matrix $\bar\rho$
for an antineutrino pure state is defined similarly.

In this paper we will assume that the $CP$-violating phase is $\delta=0$.
A brief discussion of the effect of a nonvanishing $CP$ phase is given in
Sec.~\ref{sec:conclusions}. Because we are only interested in neutrino
flavor transformation, we will assume that neutrinos are free streaming
except for forward scattering on the background medium.
For now we also will assume that there is no ordinary matter background.
For this case it is most convenient to work in the vacuum mass basis.
This basis is implicitly adopted in all the following discussions except for
Sec.~\ref{sec:matt}, where we  will discuss the effects of the 
ordinary matter potential.
To simplify the problem even further, 
we will assume that the neutrino gas is isotropic and uniform.
This corresponds
to the ``single-angle approximation'' in numerical simulations of
flavor oscillations of supernova neutrinos. It has been shown
numerically \cite{Duan:2006an,Duan:2006jv,EstebanPretel:2007ec,Fogli:2007bk} 
that both single-angle
and multi-angle calculations produce similar neutrino spectral
swaps.

For an isotropic and uniform neutrino gas confined in a fixed volume,
the equations of motion (e.o.m.) for neutrino density matrix $\rho_\bfp$ are
\begin{equation}
\rmi\dot{\rho}_\bfp=[\sfH_\bfp,\rho_\bfp].
\label{eq:eom-nu}
\end{equation}
Here the Hamiltonian for neutrino mode $\bfp$ is
\begin{equation}
\sfH_\bfp=\sfH^\vac_\bfp+
\sqrt{2}\GF\int\!\frac{\rmd^3\bfq}{(2\pi)^3}
(\rho_\bfq-\bar\rho_\bfq^*),
\label{eq:Ham}
\end{equation}
where $\GF$ is the Fermi constant, and the vacuum term in the Hamiltonian is
\begin{equation}
\sfH^\vac_\bfp=\frac{1}{2|\bfp|}\diag(m_1^2,m_2^2,m_3^2)
\end{equation} 
with $m_i^2$ being the mass-squared eigenvalues corresponding to
neutrino vacuum mass eigenstates $|\nu_i\rangle$.
For antineutrino density matrix $\bar\rho_\bfp$ one has
\begin{equation}
\rmi\dot{\bar\rho}_\bfp=[\bar{\sfH}_\bfp,\bar\rho_\bfp],
\label{eq:eom-anu}
\end{equation}
with the Hamiltonian defined as 
\begin{equation}
\bar{\sfH}_\bfp=\sfH^\vac_\bfp+
\sqrt{2}\GF\int\!\frac{\rmd^3\bfq}{(2\pi)^3}
(\bar\rho_\bfq-\rho_\bfq^*).
\label{eq:Ham-anti}
\end{equation}

We note that Ref.~\cite{Sigl:1992fn} has defined the antineutrino 
density matrix as the
complex conjugate of what one usually writes as a density matrix
(i.e., $\bar\rho_\bfp^*\rightarrow\bar\rho_\bfp$).
This notation leads to a slightly simpler version of the e.o.m.~for
both neutrinos and antineutrinos.
However, for a vanishing $CP$ phase, it is possible to treat
neutrinos and antineutrinos on an equal footing and combine
Eqs.~\eqref{eq:eom-nu} and \eqref{eq:eom-anu} into a single expression,
as  has been done in  Ref.~\cite{Duan:2005cp} 
for the two-flavor mixing scenario.
To see this, we note that  a neutrino or antineutrino mode
in an isotropic and uniform gas
is completely characterized by 
\begin{equation}
\omega(E)\equiv\mp\Tr(\sfH^\vac_\bfp\lambda_3)
=\pm\frac{\Delta m_{21}^2}{2E},
\label{eq:omega}
\end{equation}
where $E=|\bfp|$ is the energy of the neutrino or antineutrino,
 upper (lower) signs are for neutrinos (antineutrinos),
$\Delta m_{21}^2=m_2^2-m_1^2$ is approximately 
the solar mass-squared difference $\dmsol$,
and $\lambda_3$ is one of the Gell-Mann matrices
$\lambda_a$ ($a=1,\ldots,8$).
Because the number density for a neutrino (antineutrino) mode 
$\nu_\bfp$ ($\bar\nu_\bfp$) is conserved for a neutrino gas
in a fixed volume, 
we can define the total neutrino number density
\begin{equation}
n_\nu^\tot\equiv
\int\!\frac{\rmd^3\bfq}{(2\pi)^3}\Tr(\rho_\bfq+\bar\rho_\bfq),
\label{eq:ntot}
\end{equation}
and the normalized distribution function is
\begin{equation}
f_{\omega}\equiv \frac{E^2}{2\pi^2 n_\nu^\tot}
\left|\frac{\rmd E}{\rmd\omega}\right|\times
\left\{\begin{array}{ll}
\Tr(\rho_\bfp)&\text{if }\omega>0,\\
\Tr(\bar\rho_\bfp)&\text{if }\omega<0.
\end{array}\right.
\label{eq:f}
\end{equation}
Using Eqs.~\eqref{eq:ntot} and \eqref{eq:f} we can express the
integral in Eq.~\eqref{eq:eom-nu} as
\begin{equation}
\int\!\frac{\rmd^3\bfq}{(2\pi)^3}
(\rho_\bfq-\bar\rho_\bfq^*)
= n_\nu^\tot
\int_{-\infty}^\infty\!\rmd\omega f_\omega\varrho_\omega.
\label{eq:rho-varrho}
\end{equation}
In Eq.~\eqref{eq:rho-varrho} we have defined the ``reduced 
neutrino flavor density matrix'' $\varrho_\omega$ for neutrino mode $\omega$:
\begin{equation}
\varrho_{\omega}\sim\left\{\begin{array}{ll}
\rho_\bfq &\text{if }\omega>0,\\
-\bar\rho_\bfq^* &\text{if }\omega<0
\end{array}\right.
\label{eq:varrho}
\end{equation}
which has normalization
\begin{equation}
\Tr(\varrho_\omega)=\left\{\begin{array}{ll}
+1 &\text{if }\omega>0,\\
-1 &\text{if }\omega<0.
\end{array}\right.
\label{eq:varrho-norm}
\end{equation}

Using Eqs.~\eqref{eq:eom-nu}, \eqref{eq:eom-anu}
and \eqref{eq:rho-varrho} we find the e.o.m.~for $\varrho_\omega$:
\begin{equation}
\rmi\dot\varrho_\omega=[\sfH_\omega, \varrho_\omega].
\label{eq:eom}
\end{equation}
The Hamiltonian for neutrino mode $\omega$ is
\begin{equation}
\sfH_\omega=\sfH_\omega^\vac
+\mu\varrho_\tot,
\label{eq:Ham-omega}
\end{equation}
where in the vacuum mass basis
\begin{equation}
\sfH_\omega^\vac=-\omega\frac{\lambda_3}{2}-\kappa\frac{\lambda_8}{\sqrt{3}}.
\end{equation}
Here we define
\begin{equation}
\mu\equiv\sqrt{2}\GF n_\nu^\tot.
\end{equation}
This parameter dictates the strength of neutrino self-coupling.
The total neutrino flavor density matrix is defined to be
\begin{equation}
\varrho_\tot\equiv\int_{-\infty}^\infty\!\rmd\omega \,f_\omega\varrho_{\omega}.
\label{eq:varrho-tot}
\end{equation}

In Eq.~\eqref{eq:Ham-omega} 
we have left out the trace term for $\sfH_\omega$ 
(which is irrelevant for neutrino oscillations), 
and we have defined the oscillation parameter $\kappa$ to be
\begin{subequations}
\label{eq:kappa}
\begin{align}
\kappa(E)&\equiv\mp\frac{\sqrt{3}}{2}\Tr(\sfH^\vac_\bfp\lambda_8),\\
&=\pm\frac{1}{2E}\left[m_3^2-\frac{(m_1^2+m_2^2)}{2}\right],
\end{align}
\end{subequations}
where the upper (lower) signs are for neutrinos (antineutrinos).
Because the atmospheric mass-squared difference $\dmatm$
is much larger than $\dmsol$, one has
\begin{equation}
m_3^2-\frac{(m_1^2+m_2^2)}{2}\simeq\pm\dmatm,
\end{equation}
where the plus (minus) sign is for the normal (inverted)
neutrino mass hierarchy.

We note that $f_\omega$ does not change with time if there
is no inelastic scattering of neutrinos.
We also note that  Eq.~\eqref{eq:eom} is actually
more generally valid than Eqs.~\eqref{eq:eom-nu} and \eqref{eq:eom-anu} 
so long as the neutrino gas stays isotropic and uniform.
For example, this would be true for an homogeneous and isotropic
early universe, i.e., the Friedman solution. The ``single-angle
approximation'' (see, e.g., Ref.~\cite{Duan:2006an})
for flavor evolution of supernova neutrinos
is essentially equivalent to this scenario. For this case,
the flavor content of a neutrino propagating along any trajectory
at a given radius is assumed to be identical to that of a neutrino
with the same energy propagating along a radial trajectory at the same radius.
In this approximation the flavor evolution of neutrinos as a  function
of time $t$ can be represented as the flavor evolution
 of neutrinos propagating along the
radial trajectory as a function of radius $r$. In addition,
one can define the effective total neutrino number density
at each radius as
\begin{equation}
n_\nu^\tot=\frac{D(r/R_\nu)}{2\pi R_\nu^2}
\sum_{\nu}
\frac{L_\nu}{\langle E_\nu\rangle}\int_0^\infty\!\rmd E\, f_\nu(E).
\label{eq:sn-ntot}
\end{equation}
This takes  account of
both the geometric dilution and (partly) the anisotropy of
supernova neutrinos.
In Eq.~\eqref{eq:sn-ntot} $R_\nu$ is the radius of the neutrino
sphere, the geometric factor is
\begin{equation}
D(\xi)=\frac{1}{2}(1-\sqrt{1-\xi^{-2}})^2,
\end{equation}
$L_\nu$, $\langle E_\nu\rangle$ and $f_\nu(E)$
are the neutrino luminosity, average energy and normalized
energy distribution function for species $\nu$ at the neutrino sphere, 
respectively, and the summation is  over all neutrino species 
(including both neutrinos
and antineutrinos).

\subsection{Flavor Vector Description}

The flavor polarization
vector $\mathbf{P}$ \cite{Sigl:1992fn}
and neutrino flavor isospin $\mathbf{s}_\omega$ \cite{Duan:2005cp}
are important techniques that
 have been used extensively to describe the flavor mixing of neutrinos
in two-flavor mixing scenarios.
 These notations have helped in visualizing
and giving insights into the problem of collective neutrino oscillations.

To generalize these notations to the $3\times3$ case, we note that 
a $3\times3$ Hermitian matrix $\mathsf{A}$ can be expressed as 
the  linear combination of the  identity matrix $\mathsf{I}$ and 
Gell-Mann matrices $\lambda_a$:
\begin{equation}
\mathsf{A}=\frac{1}{3}\Tr(\mathsf{A})\mathsf{I}
+\sum_a A^{(a)}\frac{\lambda_a}{2},
\end{equation}
where
\begin{equation}
A^{(a)} \equiv \Tr(\mathsf{A}\lambda_a)
\end{equation}
can be viewed as the $a$'th component of an 8-dimensional vector $\mathbf{A}$.
In particular, ``flavor vector''
\begin{equation}
\bmr_\omega=(\varrho_\omega^{(1)},\ldots,\varrho_\omega^{(8)})^\mathrm{T}
\end{equation}
is the generalized version of neutrino flavor isospin $\mathbf{s}_\omega$.%
\footnote{Ref.~\cite{Dasgupta:2007ws} appears
while this manuscript was in preparation. Ref.~\cite{Dasgupta:2007ws}
has proposed a three-flavor version of the 
Bloch vector which is a generalization
of the two-flavor polarization vector defined in Ref.~\cite{Sigl:1992fn}.
The difference between the three-flavor Bloch vector and the flavor
vector defined here is similar to that between the two-flavor polarization
vector and the neutrino flavor isospin defined in Ref.~\cite{Duan:2005cp}.
In the flavor vector description,
the directions of flavor vectors and flavor isospins 
for antineutrinos are intentionally reversed
so that the e.o.m.~for flavor vectors and flavor isospins 
(for both neutrinos and antineutrinos)
can be written within a single expression. 
Likewise, the density matrix defined in Ref.~\cite{Dasgupta:2007ws}
is different from the flavor density matrix in this paper by
a sign for antineutrinos.
This is of course a notation
difference and does not affect the physical results.}

Because $\Tr(\varrho_\omega)$ is fixed by the
normalization condition in Eq.~\eqref{eq:varrho-norm},
flavor vector $\bmr_\omega$ is fully equivalent to the
density matrix $\varrho_\omega$.
In particular, the number densities of $\nu_1$, $\nu_2$ and $\nu_3$ 
in mode $\omega$
can be expressed in terms of $\varrho_\omega^{(3)}$ and $\varrho_\omega^{(8)}$:
\begin{subequations}
\label{eq:n123}
\begin{align}
n_{\nu_1}(\omega)&=n_\nu^\tot f_\omega 
\left(\frac{1}{3}
+\frac{1}{2}\varrho_\omega^{(3)}+\frac{1}{2\sqrt{3}}\varrho_\omega^{(8)}\right),\\
n_{\nu_2}(\omega)&=n_\nu^\tot f_\omega 
\left(\frac{1}{3}
-\frac{1}{2}\varrho_\omega^{(3)}+\frac{1}{2\sqrt{3}}\varrho_\omega^{(8)}\right),\\
n_{\nu_3}(\omega)&=n_\nu^\tot f_\omega 
\left(\frac{1}{3}-\frac{1}{\sqrt{3}}\varrho_\omega^{(8)}\right).
\end{align}
\end{subequations}
Noting the difference between the definition of $\varrho_\omega$
for neutrinos and antineutrinos [Eq.~\eqref{eq:varrho}], we have
\begin{subequations}
\label{eq:an123}
\begin{align}
n_{\bar\nu_1}(\omega)&=n_\nu^\tot f_\omega 
\left(\frac{1}{3}
-\frac{1}{2}\varrho_\omega^{(3)}-\frac{1}{2\sqrt{3}}\varrho_\omega^{(8)}\right),\\
n_{\bar\nu_2}(\omega)&=n_\nu^\tot f_\omega 
\left(\frac{1}{3}
+\frac{1}{2}\varrho_\omega^{(3)}-\frac{1}{2\sqrt{3}}\varrho_\omega^{(8)}\right),\\
n_{\bar\nu_3}(\omega)&=n_\nu^\tot f_\omega 
\left(\frac{1}{3}+\frac{1}{\sqrt{3}}\varrho_\omega^{(8)}\right).
\end{align}
\end{subequations}
The relation between $\omega$ and the energy of a neutrino or antineutrino
is described in Eq.~\eqref{eq:omega}.

One can define the cross  and dot products of 
two 8-dimensional vectors $\mathbf{A}$ and $\mathbf{B}$ to be
\begin{subequations}
\begin{align}
\mathbf{A}\times\mathbf{B}&\equiv
-\rmi\,\Tr([\mathsf{A},\mathsf{B}]\lambda_a)\hbe_a
=f_{abc}A^{(b)}B^{(c)}\hbe_a,
\label{eq:cross}
\\
\mathbf{A}\cdot\mathbf{B}&\equiv
2\Tr(\mathsf{A}\mathsf{B})-\frac{2}{3}\Tr(\mathsf{A})\Tr(\mathsf{B})
=A^{(a)}B^{(a)},
\end{align}
\end{subequations}
where $\hbe_a$ is the $a$'th unit vector in the
8-dimensional flavor space, and $f_{abc}$ are the antisymmetric
structure constants of SU(3):
\begin{equation}
\left[\frac{\lambda_a}{2}, \frac{\lambda_b}{2}\right]
=\rmi f_{abc}\frac{\lambda_c}{2}.
\end{equation}
The summation over Gell-Mann indices are implicitly
assumed in the above equations when they appear twice
in the subscripts or subscripts.
Using Eqs.~\eqref{eq:eom} and \eqref{eq:cross} we can write 
the e.o.m.~for flavor vector $\bmr_\omega$ as
\begin{subequations}
\label{eq:eom-vec}
\begin{align}
\frac{\rmd}{\rmd t}\bmr_\omega
&=-\bmr_\omega \times \mathbf{H}_\omega,
\label{eq:eom-vec1}\\
&=\bmr_\omega \times 
\left(\omega\hbe_3+\frac{2\kappa}{\sqrt{3}}\hbe_8\right)
-\mu\bmr_\omega\times
\int_{-\infty}^\infty\!\rmd\omega^\prime\,f_{\omega^\prime} \bmr_{\omega^\prime}.
\label{eq:eom-vec2}
\end{align}
\end{subequations}
The first term in Eq.~\eqref{eq:eom-vec2} corresponds the
 precession of $\bmr_\omega$ around an ``external field'', and the second
term corresponds a ``spin-spin anti-coupling'' with strength $\mu$.

Eq.~\eqref{eq:eom-vec1} makes it clear that the change in 
$\bmr_\omega$ is orthogonal to $\bmr_\omega$ itself, and, therefore,
the magnitude of this quantity does not changed. We have
\begin{equation}
|\bmr_\omega|^2=\sum_a[\Tr(\varrho_\omega\lambda_a)]^2=\text{const.}
\end{equation}
This is  a natural result that also follows from
the fact that Eqs.~\eqref{eq:eom-nu} and \eqref{eq:eom-anu}
maintain the coherence of $\rho_\bfp$ and $\bar\rho_\bfp$.
Following Ref.~\cite{Duan:2005cp} we define the effective energy
of the system to be
\begin{equation}
\begin{split}
\mathcal{E}&\equiv -\int_{-\infty}^\infty\!\rmd\omega\,
f_\omega\bmr_\omega \cdot 
\left(\frac{\omega}{2}\hbe_3+\frac{\kappa}{\sqrt{3}}\hbe_8\right)\\
&\quad+\frac{\mu}{4}\int_{-\infty}^\infty\!\rmd\omega
\int_{-\infty}^\infty\!\rmd\omega^\prime\,
f_\omega f_{\omega^\prime} \bmr_\omega\cdot\bmr_{\omega^\prime}.
\end{split}
\label{eq:E}
\end{equation}
Clearly the effective energy $\mathcal{E}$ of the system
is conserved if $n_\nu^\tot$ is constant.

Although flavor vector $\bmr_\omega$ seems to behave in a way similar to flavor
isospin $\mathbf{s}_\omega$ in two-flavor mixing scenarios, there
are fundamental differences between the 8-dimensional flavor
vector space and the 3-dimensional flavor isospace.
For example, two 8-dimensional vectors $\mathbf{A}$
and $\mathbf{B}$ are ``perpendicular'' or ``parallel'' to each other if 
$\mathbf{A}\cdot\mathbf{B}=0$ or $\mathbf{A}\times\mathbf{B}=0$, respectively.
Because there are two linearly independent generators of SU(3) 
that commute with each other, for any vector $\mathbf{A}$
one can always find another vector $\mathbf{A}^\prime$ which is both
``perpendicular'' and ``parallel'' to $\mathbf{A}$.
Consequently, generally speaking 
\begin{equation}
\mathbf{B}\neq\mathbf{A}\frac{\mathbf{A}\cdot\mathbf{B}}{|\mathbf{A}|^2}
\end{equation}
even if $\mathbf{B}$ is ``parallel'' to $\mathbf{A}$.

The existence of two linearly independent and commuting generators of SU(3)
has another important consequence. Because $[\lambda_3,\lambda_8]=0$,
rotations around $\hbe_3$ and $\hbe_8$ 
 can be viewed as independent. In particular,
the first term in Eq.~\eqref{eq:eom-vec2} can be interpreted as
simultaneous but independent precession of flavor vector $\bmr_\omega$
around $\hbe_3$ and $\hbe_8$ with generally
different angular velocities.

Although the density matrix and flavor vector descriptions 
are equivalent, the rotation of a flavor vector in the
8-dimensional flavor space is not easily visualizable. 
Therefore, we will base our discussions mostly on
the density matrix formalism with intermittent references
to the flavor vector notation where it seems convenient.

\subsection{Conserved ``Lepton Numbers''%
\label{sec:Ls}}

Multiplying Eq.~\eqref{eq:eom} by $f_\omega$ and integrating it over $\omega$
we obtain the e.o.m.~for $\varrho_\tot$:
\begin{equation}
\rmi\dot{\varrho}_\tot=
\frac{1}{2}
\int_{-\infty}^\infty\!\rmd\omega\,f_\omega\omega[\varrho_\omega,\lambda_3]
+\frac{1}{\sqrt{3}}
\int_{-\infty}^\infty\!\rmd\omega\,f_\omega\kappa[\varrho_\omega,\lambda_8].
\label{eq:eom-rhotot}
\end{equation}
Eq.~\eqref{eq:eom-rhotot} is not a closed equation
from which we could solve for $\varrho_\tot$.
However,
because $\lambda_3$ and $\lambda_8$
commute with each other, it is clear
that the two ``lepton numbers''
\begin{subequations}
\label{eq:Ls}
\begin{align}
L_3&\equiv\varrho_\tot^{(3)}=\int_{-\infty}^\infty\!\rmd\omega\,f_\omega
\Tr(\varrho_\omega\lambda_3)\\
\intertext{and}
L_8&\equiv\varrho_\tot^{(8)}=\int_{-\infty}^\infty\!\rmd\omega\,f_\omega
\Tr(\varrho_\omega\lambda_8)
\end{align}
\end{subequations}
are constants of the motion.
Because $\Tr(\varrho_\tot)$ does not change with time,
 the lepton number (fraction) in each vacuum
mass eigenstate is individually conserved.

\subsection{Corotating Frame%
\label{sec:cr-frame}}

In the density matrix description, changing from the
static frame to a corotating frame 
corresponds to a transformation
\begin{equation}
\varrho_\omega\rightarrow 
\tilde\varrho_\omega\equiv e^{\rmi\Hcr t}
\varrho_\omega e^{-\rmi\Hcr t},
\end{equation}
where $\Hcr$ is a $3\times3$ Hermitian matrix 
that is common to all neutrino modes and does not change with time.
The flavor density matrix $\tilde\varrho_\omega$
in the corotating frame satisfies an e.o.m.~similar to
that in Eq.~\eqref{eq:eom}:
\begin{equation}
\rmi\dot{\tilde\varrho}_\omega=[\tilde{\sfH}_\omega, \tilde\varrho_\omega],
\end{equation}
where 
\begin{equation}
\tilde{\sfH}_\omega\equiv e^{\rmi\Hcr t}\sfH_\omega e^{-\rmi\Hcr t}-\Hcr
\end{equation}
is the Hamiltonian for neutrino mode $\omega$ in the corotating
frame associated with $\Hcr$.

A very special set of corotating frames corresponds to
simultaneous rotations around $\hbe_3$ and $\hbe_8$ with
angular velocities $\Omega$ and $2K/\sqrt{3}$, respectively, and
\begin{equation}
\Hcr=-\Omega\frac{\lambda_3}{2}-K\frac{\lambda_8}{\sqrt{3}}.
\end{equation}
Because $\sfH_\omega^\vac$, $\lambda_3$
and $\lambda_8$ commute with each other,
 $\tilde{\sfH}_\omega$ in these special corotating frames 
takes the same form as $\sfH_\omega$
in Eq.~\eqref{eq:Ham-omega}
except for the replacements
\begin{subequations}
\begin{align}
\omega&\rightarrow\omega-\Omega,\\
\kappa&\rightarrow\kappa-K,\\
\varrho_\tot&\rightarrow\tilde\varrho_\tot
=e^{\rmi\Hcr t}\varrho_\tot e^{-\rmi\Hcr t}.
\end{align}
\end{subequations}
Also, because the occupation numbers of each 
vacuum eigenstate in a neutrino mode $\omega$ are
determined only by $\varrho_\omega^{(3)}$ and $\varrho_\omega^{(8)}$,
 the probability for the neutrino mode
to be in the $i$'th vacuum mass eigenstate in these special
corotating frames, $|\tilde{a}_{\nu_i}(\omega)|^2$, 
is the same as  
the probability for the neutrino mode to be in the same eigenstate
in the static frame, $|a_{\nu_i}(\omega)|^2$.
Therefore, lepton numbers $L_3$ and $L_8$ are not
changed by the corotating-frame transformation either.

\section{Adiabatic/Precession Solution 
and Stepwise Spectral Swapping%
\label{sec:sol}}

We now seek  a natural extension to the adiabatic/precession solution
presented in Ref.~\cite{Raffelt:2007cb}.
This solution, like the $2\times2$ case,
will be essentially a quasi-static solution that, for given $n_\nu^\tot$,
 is the same as the ``static'' solution  which satisfies the
\textit{precession ansatz}. There is a family of static solutions
for each value of $n_\nu^\tot$. So long as $n_\nu^\tot$ changes slowly,
a particular solution in each family parametrized by $n_\nu^\tot$
is uniquely determined by the initial conditions and by the
\textit{adiabatic ansatz}. This gives the adiabatic/precession solution.
We will discuss the precession ansatz and  three-flavor
synchronization in Sec.~\ref{sec:prec}. 
 In Sec.~\ref{sec:adiabatic} we will  discuss
the adiabatic ansatz and outline a formal procedure for
obtaining a three-flavor
adiabatic/precession solution.
In Sec.~\ref{sec:swapping} we will illustrate with a
numerical example how  stepwise 
spectral swapping arises from the three-flavor
adiabatic/precession solution  as $n_\nu^\tot\rightarrow0$.

\subsection{The Precession Ansatz and Synchronization%
\label{sec:prec}}

The \textit{precession ansatz} is that, for constant $n_\nu^\tot$, 
 it is possible to find a Hermitian matrix
\begin{equation}
\Hcr=-\wpr\frac{\lambda_3}{2}-\kpr\frac{\lambda_8}{\sqrt{3}}
\end{equation}
such that the flavor density matrix $\tilde\varrho_\omega$
is static in the corotating frame associated with $\Hcr$, i.e.
\begin{equation}
[\tilde\sfH_\omega,\tilde\varrho_\omega] = 0.
\label{eq:prec-ansatz}
\end{equation}
From the arguments in Sec.~\ref{sec:cr-frame},
if the precession ansatz is satisfied,
$|a_{\nu_i}(\omega)|^2$ does not change with time.
In this sense, a solution that satisfies the precession
ansatz is ``static''.

When the adiabatic ansatz is satisfied,
each flavor vector 
$\bmr_\omega$ will  precess uniformly around
$\hbe_3$ and $\hbe_8$.  In other words, the system
is in a state that is symmetric about $\hbe_3$ and $\hbe_8$.
Because the e.o.m.~for $\bmr_\omega$ [Eqs.~\eqref{eq:eom-vec}] possess 
the same symmetry around the $\hbe_3$ and $\hbe_8$ axes,
we expect ``static'' precession solutions
to exist for constant $n_\nu^\tot$ 
with appropriate initial conditions. 
In particular, this symmetry obtains approximately in 
dense neutrino gases where
\begin{equation}
\mu=\sqrt{2}\GF n_\nu^\tot\gg|\kappa|\gg|\omega|
\label{eq:large-n}
\end{equation}
for most neutrino modes.
This conclusion can be shown as follows. 

When Eq.~\eqref{eq:large-n} is
satisfied, the conserved effective energy of the system is 
[Eq.~\eqref{eq:E}]
\begin{equation}
\mathcal{E}\simeq\frac{\mu}{4}|\bmr_\tot|^2
\end{equation}
and, therefore,
\begin{equation}
|\bmr_\tot|\simeq\text{const.}
\end{equation}
Because the two lepton numbers $L_3=\varrho_\tot^{(3)}$
and $L_8=\varrho_\tot^{(8)}$ are conserved (Sec.~\ref{sec:Ls}),
the total flavor vector $\bmr_\tot$ must  precess simultaneously
around $\hbe_3$ and $\hbe_8$ according to
\begin{equation}
\frac{\rmd}{\rmd t}\bmr_\tot\simeq
\bmr_\tot\times\left(\wpr^\infty\hbe_3+\frac{2\kpr^\infty}{\sqrt{3}}\hbe_8\right).
\label{eq:eom-ln-tot}
\end{equation}
Also in this limit, Eq.~\eqref{eq:eom-vec} becomes
\begin{equation}
\frac{\rmd}{\rmd t}\bmr_\omega\simeq
-\mu\bmr_\omega\times\bmr_\tot.
\label{eq:eom-ln}
\end{equation}

Eqs.~\eqref{eq:eom-ln-tot} and \eqref{eq:eom-ln} suggests a
simple geometric picture for the flavor evolution in dense
neutrino gases. On short time scales ($\Delta t\sim\mu^{-1}$),
each flavor vector $\bmr_\omega$  precess rapidly around the 
total flavor vector $\bmr_\tot$. On large time scales 
($\Delta t\sim|\wpr^\infty|^{-1},|\kpr^\infty|^{-1}$),
all flavor vectors precess slowly around $\hbe_3$
and $\hbe_8$ with angular velocities $\wpr^\infty$ and $2\kpr^\infty/\sqrt{3}$,
respectively. This is analogous  to the synchronization phenomenon
in two-flavor mixing scenarios \cite{Pastor:2001iu}.

\subsection{The Adiabatic Ansatz and Adiabatic/Precession Solutions%
\label{sec:adiabatic}}

When the precession ansatz in Eq.~\eqref{eq:prec-ansatz}
is satisfied, it is possible to find a unitary matrix $\sfX_\omega$
that simultaneously diagonalizes both
$\tilde{\sfH}_\omega$ and $\tilde\varrho_\omega$:
\begin{subequations}
\begin{align}
\sfX_\omega\tilde{\sfH}_\omega\sfX_\omega^\dagger
&=\diag(\wtm{L},\wtm{M},\wtm{H}),
\label{eq:diag-H}\\
\sfX_\omega\tilde\varrho_\omega\sfX_\omega^\dagger
&=\pm\diag(|a_{\ntm{L}}|^2,|a_{\ntm{M}}|^2,|a_{\ntm{H}}|^2),
\label{eq:diag-varrho}
\end{align}
\end{subequations}
where $\wtm{L}<\wtm{M}<\wtm{H}$
are the eigenvalues corresponding to
 the eigenstates of $\tilde{\sfH}_\omega$,
$|\ntm{L}(\omega)\rangle$, $|\ntm{M}(\omega)\rangle$ 
and $|\ntm{H}(\omega)\rangle$.
In Eq.~\eqref{eq:diag-varrho} 
the plus (minus) sign is for neutrinos (antineutrinos).
The \textit{adiabatic ansatz} is simply that 
\begin{equation}
|a_{\tilde\nu_l}(\omega)|^2=\text{const.}\quad(l=\text{L, M, H})
\label{eq:adiabatic-ansatz}
\end{equation}
 as $n_\nu^\tot$ slowly varies with time.

As discussed in Ref.~\cite{Duan:2007fw}, this ``adiabaticity''
criterion connects neutrino systems in different corotating frames
at different values of $n_\nu^\tot$.
Note that this adiabaticity criterion 
is different from the meaning of adiabaticity usually adopted in
the literature, e.g., when discussing the MSW mechanism,
which is always based on the static frame. 
Following Ref.~\cite{Raffelt:2007cb}, we argue here that the
adiabatic ansatz can be satisfied if, for each 
neutrino mode $\omega$,
$\tilde{\bfH}_\omega$ rotates at a speed much slower
than precession rate of $\tilde{\bmr}_\omega$ around $\tilde{\bfH}_\omega$,
i.e.
\begin{equation}
\gamma\equiv
\frac{|\tilde{\bfH}_\omega\times\rmd \tilde{\bfH}_\omega/\rmd t|}%
{|\tilde{\bfH}_\omega|^3}\ll1.
\label{eq:adiabatic}
\end{equation}

The adiabatic/precession solution can be obtained formally  by 
employing the following procedure:\begin{enumerate}
\item At any $n_\nu^\tot$ find for each neutrino mode $\omega$ 
a unitary matrix $\sfX_\omega$
that diagonalizes $\tilde{\sfH}_\omega$. Matrix $\sfX_\omega$
is expressed as a function of $(\tilde\varrho_\tot,\wpr,\kpr)$.
\item\label{step:varrho-omega} 
For given initial values of $|a_{\tilde\nu_l}(\omega)|^2$, find 
$\tilde\varrho_\omega=
\pm\sfX_\omega^\dagger
\diag(|a_{\ntm{L}}|^2,|a_{\ntm{M}}|^2,|a_{\ntm{H}}|^2)\sfX_\omega$
as a function of $(\tilde\varrho_\tot,\wpr,\kpr)$.
\item For given initial lepton numbers $L_3$ and $L_8$
 find $\tilde\varrho_\tot^{(3)}=L_3$ and
$\tilde\varrho_\tot^{(8)}=L_8$. 
From the definition 
$\tilde\varrho_\tot=\int_{-\infty}^\infty\!\rmd\omega f_\omega
\tilde\varrho_\omega(\tilde\varrho_\tot,\wpr,\kpr)$, solve for the
precession angular velocities
$\wpr$ and $\kpr$ and the remaining components of $\tilde\varrho_\tot$.
\item Find $\tilde\varrho_\omega$ for each neutrino mode $\omega$
using the expression $\tilde\varrho_\omega(\tilde\varrho_\tot,\wpr,\kpr)$
obtained in step \ref{step:varrho-omega}.
\end{enumerate}

We note that the above procedure actually gives a set of equivalent
solutions. This is because the precession solution is symmetric
around $\hbe_3$ and $\hbe_8$. If $(\tilde\varrho_\tot,\wpr,\kpr)$
is a solution, $(\tilde\varrho_\tot^\prime,\wpr,\kpr)$ is also a solution,
where $\tilde\varrho_\tot^\prime$ is related to $\tilde\varrho_\tot$
by two arbitrary phases $\phi_3$ and $\phi_8$:
\begin{equation}
\tilde{\varrho}_\tot^\prime=
\exp\left(-\rmi\phi_3\frac{\lambda_3}{2}-\rmi\phi_8\frac{\lambda_8}{2}\right)
\tilde{\varrho}_\tot
\exp\left(\rmi\phi_3\frac{\lambda_3}{2}+\rmi\phi_8\frac{\lambda_8}{2}\right).
\end{equation}
One can fix these two phases by, e.g., choosing 
$\tilde{\varrho}_\tot^{(2)}=\tilde{\varrho}_\tot^{(7)}=0$.

\subsection{Stepwise Spectral Swapping%
\label{sec:swapping}}

Neutrino flavor mixing becomes very simple 
in the adiabatic/precession solution
presented in Sec.~\ref{sec:adiabatic} when $n_\nu^\tot\rightarrow0$.
In this limit $\tilde{\sfH}_\omega$ is diagonal in the vacuum
mass basis:
\begin{equation}
\tilde{\sfH}_\omega|_{n_\nu^\tot\rightarrow0}
=\diag(\tilde\omega_1,\tilde\omega_2,
\tilde\omega_3)_{n_\nu^\tot\rightarrow0},
\label{eq:H-cr-0n}
\end{equation}
where
\begin{subequations}
\label{eq:omega123-0n}
\begin{align}
\tilde\omega_1|_{n_\nu^\tot\rightarrow0}
&=-\frac{1}{2}(\omega-\wpr^0)-\frac{1}{3}(\kappa-\kpr^0),\\
\tilde\omega_2|_{n_\nu^\tot\rightarrow0}
&=\frac{1}{2}(\omega-\wpr^0)-\frac{1}{3}(\kappa-\kpr^0),\\
\tilde\omega_3|_{n_\nu^\tot\rightarrow0}
&=\frac{2}{3}(\kappa-\kpr^0),
\end{align}
\end{subequations}
with $\wpr^0$ and $\kpr^0$ being the collective precession angular
velocities as $n_\nu^\tot\rightarrow0$. Eq.~\eqref{eq:H-cr-0n}
shows that $\tilde{\sfH}_\omega|_{n_\nu^\tot\rightarrow0}$ has
3 critical values of $\omega$ at which any two of its eigenvalues are equal.
These critical values are
\begin{subequations}
\begin{align}
\wc_1&=\wpr^0,\\
\wc_2&=\frac{\Delta m_{21}^2}{\Delta m_{31}^2}
\left(\kpr^0+\frac{1}{2}\wpr^0\right),\\
\wc_3&=\frac{\Delta m_{21}^2}{\Delta m_{32}^2}
\left(\kpr^0-\frac{1}{2}\wpr^0\right).
\end{align}
\end{subequations}
In practice, however, two of the critical points are usually
indistinguishable 
\begin{subequations}
\begin{align}
\wc_1&=\wcsol\equiv\wpr^0,\\
\wc_2&\simeq\wc_3\simeq\wcatm\equiv\pm\frac{\dmsol}{\dmatm}\kpr^0,
\label{eq:watm}
\end{align}
\end{subequations}
because $\dmatm\gg\dmsol$ and, therefore, $|\kpr^0|\gg|\wpr^0|$.
In Eq.~\eqref{eq:watm} 
the positive (negative) sign is for the normal (inverted)
neutrino mass hierarchy.
A critical value $\wc$ corresponds to the energy  where
``stepwise spectral swapping'' or a ``spectral split'' occurs.
If $\wc>0$, the spectral swap occurs at neutrino energy
\begin{equation}
E^\mathrm{s}\simeq\frac{\dmsol}{2|\wc|}
\end{equation}
in the neutrino sector. 
If $\wc<0$, the swap is located  in the antineutrino sector 
at energy $E^\mathrm{s}$.

We can illustrate stepwise spectral swapping by
using the following test example.
We assume a bare, hot, spherical neutron-star that
isotropically emits neutrinos directly into the vacuum from its
infinitely thin neutrino sphere. We adopt the
single-angle approximation and take the radius 
of the neutrino sphere to be $R_\nu=30$ km, and the luminosity for
each neutrino species to be 
$L_\nu=10^{52}\,\mathrm{erg/s}$.
We take the energy spectra of neutrinos to be
of the Fermi-Dirac form:
\begin{equation}
f_\nu(E)=\frac{1}{T_\nu^3 F_2(\eta_\nu)}
\frac{E^2}{\exp(E/T_\nu-\eta_\nu)+1},
\end{equation}
where 
\begin{equation}
F_k(\eta)=\int_0^\infty\frac{x^k\rmd x}{\exp(x-\eta)+1}.
\end{equation}
We take degeneracy parameters to be the same for all
neutrino species, $\eta_\nu=3$, and we choose $T_\nu$ to be such
that the average energies for various neutrino species are
$\langle E_{\nu_e}\rangle=11$ MeV, $\langle E_{\bar\nu_e}\rangle=16$ MeV
and $\langle E_{\nu_\mu}\rangle=\langle E_{\bar\nu_\mu}\rangle
=\langle E_{\nu_\tau}\rangle=\langle E_{\bar\nu_\tau}\rangle=25$ MeV,  
respectively.
For the neutrino mixing parameters 
(see, e.g., Ref.~\cite{PDBook} for our conventions)
we take $\theta_{12}=0.6$,
$\theta_{13}=0.1$, $\theta_{23}=\pi/4$, $\delta=0$,
$\Delta m_{21}^2=8\times10^{-5}\,\mathrm{eV}^2$ and
$\Delta m_{32}^2=-3\times10^{-3}\,\mathrm{eV}^2$
(inverted neutrino mass hierarchy). 

\begin{figure*}
\begin{center}
\includegraphics*[width=\textwidth, keepaspectratio]{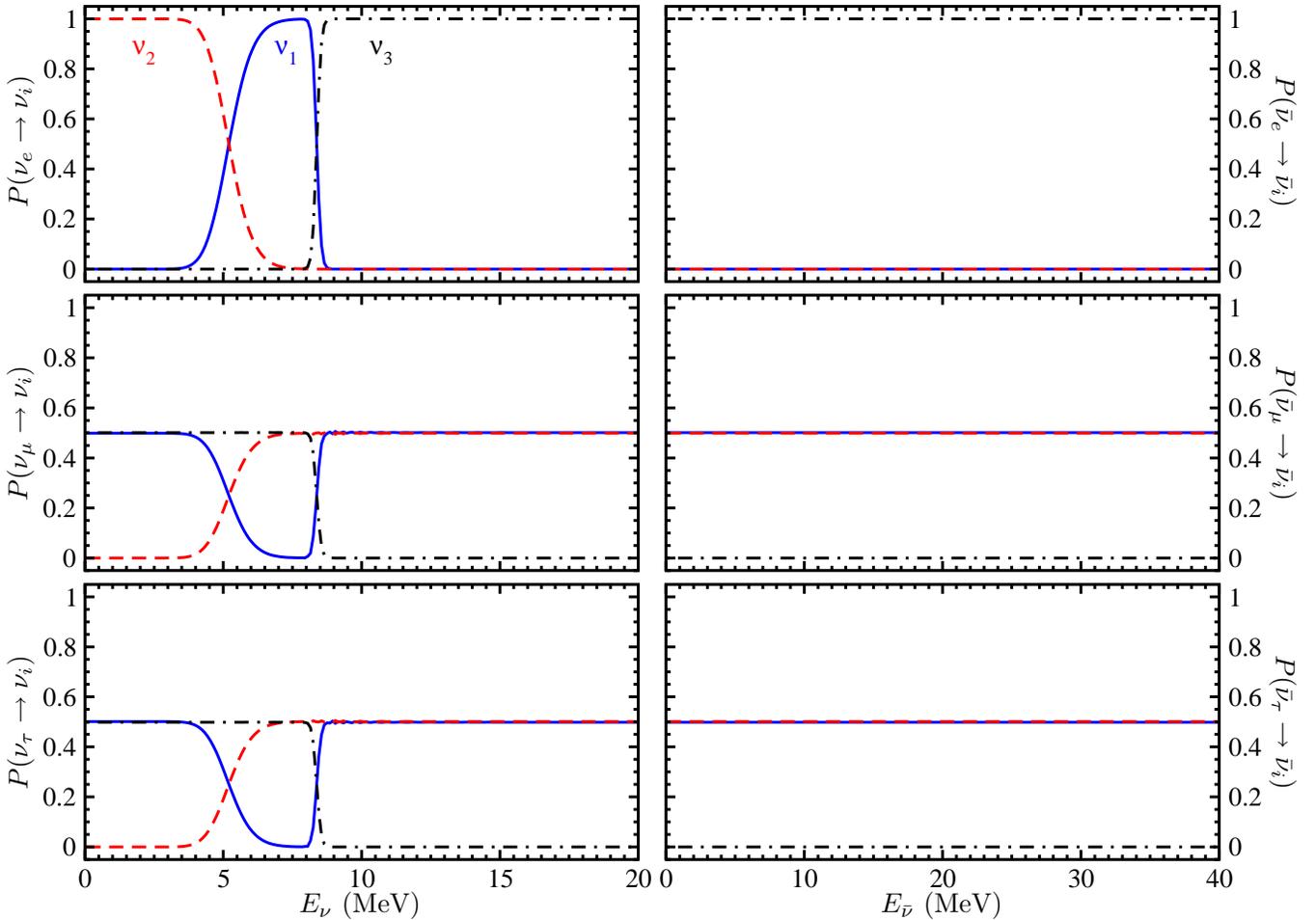} 
\end{center}
\caption{\label{fig:swap}(Color online)
Conversion probabilities $P(\nu_\alpha\rightarrow\nu_i)$ (left panels)
and $P(\bar\nu_\alpha\rightarrow\bar\nu_i)$ (right panels) as functions of 
neutrino and antineutrino energies $E_\nu$ and $E_{\bar\nu}$,
respectively, in the bare, hot neutron star example. 
The top, middle and bottom panels are for neutrinos that are
initially in pure $e$, $\mu$ and $\tau$ flavors, respectively.
The solid, dashed and dot-dashed lines are for neutrinos
that end up in the 1st, 2nd and 3rd vacuum mass eigenstates, respectively,
when $n_\nu^\tot\rightarrow0$.}
\end{figure*}

We define $P(\nu_\alpha\rightarrow\nu_i)$ and 
$P(\bar\nu_\alpha\rightarrow\bar\nu_i)$ as
the probabilities for neutrinos and antineutrinos that are
initially in pure $\alpha$ flavor state at the neutrino sphere
to   end up
in the $i$'th vacuum mass eigenstate
as $n_\nu^\tot\rightarrow0$.
In Fig.~\ref{fig:swap} we show neutrino conversion
probabilities $P(\nu_\alpha\rightarrow\nu_i)$ and 
$P(\bar\nu_\alpha\rightarrow\bar\nu_i)$ 
 as functions
of neutrino energies in our test example.
 We observe
that while $P(\nu_\alpha\rightarrow\nu_i)$ shows two swaps at
$\Ecsol\simeq5.2$ MeV and $\Ecatm\simeq8.4$ MeV, respectively,
$P(\bar\nu_\alpha\rightarrow\bar\nu_i)$ shows no swap at all. This
phenomenon can be explained using the adiabatic/precession 
solution discussed above.

\begin{table}[ht]
\caption{\label{tab:nu-ini}The correspondence between neutrino states
$|\tilde\nu_l(\omega)\rangle$ and $|\nu_\alpha\rangle$ or 
$|\bar\nu_\alpha\rangle$
at $r=R_\nu$ for the inverted neutrino
mass hierarchy case with mixing angles
$\theta_{13}\simeq0$ and $\theta_{23}\simeq\pi/4$.}
\begin{ruledtabular}
\begin{tabular}{c|c|c|c}
& $\omega<0$ & $0<\omega<\omega_\mathrm{atm}^\mathrm{sync}$
& $\omega>\omega_\mathrm{atm}^\mathrm{sync}$\\
\hline
$|\ntm{L}(\omega)\rangle$
& $\frac{1}{\sqrt{2}}(|\bar\nu_\mu\rangle-|\bar\nu_\tau\rangle)$
& $\frac{1}{\sqrt{2}}(|\nu_\mu\rangle-|\nu_\tau\rangle)$
& $\frac{1}{\sqrt{2}}(|\nu_\mu\rangle+|\nu_\tau\rangle)$\\
$|\ntm{M}(\omega)\rangle$
& $\frac{1}{\sqrt{2}}(|\bar\nu_\mu\rangle+|\bar\nu_\tau\rangle)$
& $\frac{1}{\sqrt{2}}(|\nu_\mu\rangle+|\nu_\tau\rangle)$
& $\frac{1}{\sqrt{2}}(|\nu_\mu\rangle-|\nu_\tau\rangle)$\\
$|\ntm{H}(\omega)\rangle$& $|\bar\nu_e\rangle$ 
& $|\nu_e\rangle$ & $|\nu_e\rangle$
\end{tabular}
\end{ruledtabular}
\end{table}

\begin{table}[ht]
\caption{\label{tab:nu-fin}The correspondence between neutrino states
$|\tilde\nu_l(\omega)\rangle$ and $|\nu_i\rangle$ or $|\bar\nu_i\rangle$
as $n_\nu^\tot\rightarrow0$ for the 
inverted mass hierarchy case and for swap points
with hierarchy $\wcsol>\wcatm>0$.}
\begin{ruledtabular}
\begin{tabular}{c|c|c|c|c}
& $\omega<0$ & $0<\omega<\wcatm$ & $\wcatm<\omega<\wcsol$ & $\omega>\wcsol$\\
\hline
$|\ntm{L}(\omega)\rangle$& $|\bar\nu_2\rangle$ & $|\nu_2\rangle$ 
& $|\nu_3\rangle$ & $|\nu_3\rangle$ \\
$|\ntm{M}(\omega)\rangle$& $|\bar\nu_1\rangle$ & $|\nu_1\rangle$  
& $|\nu_2\rangle$ & $|\nu_1\rangle$ \\
$|\ntm{H}(\omega)\rangle$& $|\bar\nu_3\rangle$&  $|\nu_3\rangle$ 
& $|\nu_1\rangle$  & $|\nu_2\rangle$ 
\end{tabular}
\end{ruledtabular}
\end{table}

\begin{table}[ht]
\caption{\label{tab:P} Nonzero neutrino conversion probabilities 
$P(\nu_\alpha\rightarrow\nu_i)$ and $P(\bar\nu_\alpha\rightarrow\bar\nu_i)$
based on Tables \ref{tab:nu-ini} and \ref{tab:nu-fin}
and the adiabaticity ansatz in Eq.~\eqref{eq:equal-as}.}
\begin{ruledtabular}
\begin{tabular}{c|c}
energy range & nonzero conversion probabilities\\
\hline
any $E_{\bar\nu}$ & 
$P(\bar\nu_e\rightarrow\bar\nu_3)=
2P(\bar\nu_{\mu,\tau}\rightarrow\bar\nu_{1,2})=1$\\
$E_\nu>\Ecatm$ &
$P(\nu_e\rightarrow\nu_3)=
2P(\nu_{\mu,\tau}\rightarrow\nu_{1,2})=1$\\
$\Ecsol<E_\nu<\Ecatm$ &
 $P(\nu_e\rightarrow\nu_1)=
2P(\nu_{\mu,\tau}\rightarrow\nu_{2,3})=1$\\
$E_\nu<\Ecsol$ &
 $P(\nu_e\rightarrow\nu_2)=
2P(\nu_{\mu,\tau}\rightarrow\nu_{1,3})=1$
\end{tabular}
\end{ruledtabular}
\end{table}

Assuming that the precession ansatz in Eq.~\eqref{eq:prec-ansatz} is satisfied,
we can diagonalize the Hamiltonian $\tilde{\sfH}_\omega|_{r=R_\nu}$
in the corotating frame in which all flavor vectors are static.
We denote $\wtm{L}<\wtm{M}<\wtm{H}$ as eigenvalues corresponding to the
eigenstates $|\ntm{L}\rangle$, $|\ntm{M}\rangle$ and $|\ntm{H}\rangle$
of $\tilde{\sfH}_\omega|_{r=R_\nu}$, respectively.
Because $n_\nu^\tot$ is very large at the neutrino sphere, we have
\begin{equation}
\sqrt{2}\GF (n_{\nu_e}-n_{\bar\nu_e}) \gg |\omega-\wpr^\infty|, |\kappa-\kpr^\infty|
\end{equation}
for most neutrino modes. Therefore, 
\begin{equation}
|\ntm{H}(\omega)\rangle_{r=R_\nu}\simeq\left\{\begin{array}{ll}
|\nu_e\rangle& \text{ if } \omega>0,\\
|\bar\nu_e\rangle& \text{ if } \omega<0.
\end{array}\right.
\end{equation}
We note that $\wpr$ and $\kpr$ are essentially a kind of average
of $\omega$ and $\kappa$ in the system. In our example, neutrinos
(instead of antineutrinos) are the dominant species and
the neutrino mass hierarchy is inverted. So we expect
$\wpr>0>\kpr$ at any value of $n_\nu^\tot$.
We can diagonalize the $\mu\tau$-submatrix of $\tilde{\sfH}_\omega^\vac$
in the flavor basis and obtain
\begin{subequations}
\begin{align}
|\ntm{L}(\omega)\rangle_{r=R_\nu}&\simeq
\frac{1}{\sqrt{2}}(|\bar\nu_\mu\rangle-|\bar\nu_\tau\rangle),\\
|\ntm{M}(\omega)\rangle_{r=R_\nu}&\simeq
\frac{1}{\sqrt{2}}(|\bar\nu_\mu\rangle+|\bar\nu_\tau\rangle),
\end{align}
\end{subequations}
if $\omega<0$, and
\begin{subequations}
\label{eq:ntm-Rn}
\begin{align}
|\ntm{L}(\omega)\rangle_{r=R_\nu}&\simeq
\frac{1}{\sqrt{2}}(|\nu_\mu\rangle\mp|\nu_\tau\rangle),\\
|\ntm{M}(\omega)\rangle_{r=R_\nu}&\simeq
\frac{1}{\sqrt{2}}(|\nu_\mu\rangle\pm|\nu_\tau\rangle),
\end{align}
\end{subequations}
if $\omega>0$. In Eq.~\eqref{eq:ntm-Rn},
 the upper and lower signs are for the cases
where $\omega$ is smaller or larger than
\begin{equation}
\omega_\mathrm{atm}^\mathrm{sync}=-\kpr^\infty\frac{\dmsol}{\dmatm},
\end{equation}
respectively.
Here $|\ntm{L}(\omega)\rangle_{r=R_\nu}$ and 
$|\ntm{M}(\omega)\rangle_{r=R_\nu}$
are  ``equal mixes'' of $|\nu_\mu\rangle$ and $|\nu_\tau\rangle$
or $|\bar\nu_\mu\rangle$ and $|\bar\nu_\tau\rangle$. These
results are summarized in
 Table \ref{tab:nu-ini}.

Far away from the neutron star where $n_\nu^\tot\rightarrow0$, we obtain
$|\tilde\nu_l\rangle_{n_\nu^\tot\rightarrow0}$ using 
Eqs.~\eqref{eq:H-cr-0n} and \eqref{eq:omega123-0n}.
For $\omega<0$ (antineutrinos) we have
\begin{subequations}
\begin{align}
|\ntm{L}(\omega)\rangle_{n_\nu^\tot\rightarrow0}&=|\bar\nu_2\rangle,\\
|\ntm{M}(\omega)\rangle_{n_\nu^\tot\rightarrow0}&=|\bar\nu_1\rangle,\\
|\ntm{H}(\omega)\rangle_{n_\nu^\tot\rightarrow0}&=|\bar\nu_3\rangle.
\end{align}
\end{subequations}
Assuming that $\wcsol>\wcatm>0$, we summarize the correspondence
between $|\tilde\nu_l(\omega)\rangle$ and $|\nu_i\rangle$
for $n_\nu^\tot\rightarrow0$ in Table \ref{tab:nu-fin}.

If the adiabatic ansatz in Eq.~\eqref{eq:adiabatic-ansatz} is satisfied,
we have 
\begin{equation}
|a_{\tilde\nu_l}(\omega)|^2_{r=R_\nu}
=|a_{\tilde\nu_l}(\omega)|^2_{n_\nu^\tot\rightarrow0}
\label{eq:equal-as}
\end{equation}
for each neutrino mode $\omega$. Using Eq.~\eqref{eq:equal-as}
and  Tables \ref{tab:nu-ini} and \ref{tab:nu-fin}, we can obtain
the neutrino conversion probabilities 
$P(\nu_\alpha\rightarrow\nu_i)$ and $P(\bar\nu_\alpha\rightarrow\bar\nu_i)$
for each neutrino mode $\omega$.
In Table \ref{tab:P} we summarize the values of
nonzero neutrino conversion probability
when the flavor evolution of the neutrino gas follows the 
adiabatic/precession solution.
The results in Table \ref{tab:P} and Fig.~\ref{fig:swap} 
are in good agreement.
The neutrino spectral swapping feature in
Fig.~\ref{fig:swap} indeed can be explained by the adiabatic/precession
solution.

As in two-flavor scenarios, the swapping energies $\Ecsol$ and $\Ecatm$
can be obtained by using the conservation of lepton numbers
and by assuming that
the spectral swaps are infinitely sharp in energy.
From conservation of $L_8=\varrho_\tot^{(8)}$ we have
\begin{widetext}
\begin{subequations}
\label{eq:L8-Ecatm}
\begin{eqnarray}
\sqrt{3}L_8&\stackrel{r=R_\nu}{\simeq}&
\int_0^\infty\!\rmd E\,
(\tilf_{\nu_e}-\tilf_{\bar\nu_e}),\\
&\stackrel{n_\nu^\tot\rightarrow0}{\simeq}&
\int_0^{\Ecatm}\!\rmd E\,(\tilf_{\nu_e}-\tilf_{\nu_x})
+\int_{\Ecatm}^\infty\!\rmd E\,(2\tilf_{\nu_x}-2\tilf_{\nu_e})
-\int_0^\infty\!\rmd E\,(2\tilf_{\bar\nu_x}-2\tilf_{\bar\nu_e}),
\end{eqnarray}
\end{subequations}
\end{widetext}
where $\tilf_\nu(E)$ is a distribution function which satisfies
\begin{equation}
\tilf_\nu(E)\propto f_\nu(E)
\end{equation}  
with  normalization condition
\begin{equation}
\sum_\nu\int_0^\infty\!\rmd E\,\tilf_\nu(E)=1,
\end{equation} 
and
\begin{equation}
\tilf_{\nu_x}=\tilf_{\nu_\mu}=\tilf_{\bar\nu_\mu}
=\tilf_{\nu_\tau}=\tilf_{\bar\nu_\tau}.
\end{equation}
From Eq.~\eqref{eq:L8-Ecatm} we can determine that $\Ecatm\simeq8.4$ MeV. 
We also can obtain $\Ecsol$ from the conservation of $L_3=\varrho_\tot^{(3)}$:
\begin{subequations}
\label{eq:L3-Ecsol}
\begin{eqnarray}
L_3&\stackrel{r=R_\nu}{\simeq}&
\int_0^\infty\!\rmd E\,
\cos2\theta_{12}(\tilf_{\nu_e}-\tilf_{\bar\nu_e}),\\
&\stackrel{n_\nu^\tot\rightarrow0}{\simeq}&
\int_0^{\Ecsol}\!\rmd E\,(\tilf_{\nu_x}-\tilf_{\nu_e})\nonumber\\
&&+\int_{\Ecsol}^{\Ecatm}\!\rmd E\,(\tilf_{\nu_e}-\tilf_{\nu_x}).
\end{eqnarray}
\end{subequations}
This implies $\Ecsol\simeq5.3$ MeV.
The  values for $\Ecatm$ and $\Ecsol$ derived from
the adiabatic/precession solution are also in good
agreement with the numerical results
 shown in Fig.~\ref{fig:swap}.


We  note that, in this numerical example, neutrinos of the $\mu$
and $\tau$ flavors are equally mixed and have identical energy
spectra initially. In this particular case,
the two-flavor approximation
 with effective mixing parameters
$\Delta m^2\simeq-\dmatm$ and $\theta\simeq\theta_{13}$  produce
a similar spectral swapping feature around $\Ec=\Ecatm$.
In the two-flavor mixing scenario,
the value for $\Ec$ is also determined by the conservation
of a lepton number
\begin{subequations}
\label{eq:L-2x2}
\begin{eqnarray}
L_{2\times2}&\stackrel{r=R_\nu}{\simeq}
&I_1+I_2,\\
&\stackrel{n_\nu^\tot\rightarrow0}{\simeq}
&I_1-I_2,
\end{eqnarray}
\end{subequations}
where
\begin{subequations}
\begin{align}
I_1&=\int_0^{\Ec}\!\rmd E\,(\tilf_{\nu_e}-\tilf_{\nu_x}),\\
I_2&=\int_{\Ec}^\infty\!\rmd E\,(\tilf_{\nu_e}-\tilf_{\nu_x})
-\int_0^\infty\!\rmd E\,(\tilf_{\bar\nu_e}-\tilf_{\nu_x}).
\end{align}
\end{subequations}
From Eq.~\eqref{eq:L-2x2} it is easy to see that the integral $I_2=0$. 
Comparing Eqs.~\eqref{eq:L8-Ecatm}
and \eqref{eq:L-2x2}, we see that a solution for $\Ec$ in Eq.~\eqref{eq:L-2x2}
is a solution for $\Ecatm=\Ec$ in Eq.~\eqref{eq:L8-Ecatm}.
Therefore, the two-flavor spectral swap phenomenon is completely
consistent  with the
three-flavor spectral swap phenomenon  at the atmospheric 
mass-squared-difference scale.

We also note that, in this example, there would exist only one
spectral swap at energy $E_\nu\simeq\Ecatm$ if we had
chosen the other hierarchy for swap points, i.e.
$\wcatm>\wcsol>0$. It is generally not
possible for a single spectral swap to satisfy the conservation
of both lepton numbers and, therefore, this hierarchy for
swap points is not physical.


The example given here can be generalized into
a generic procedure to predict neutrino spectral swaps:\begin{enumerate}
\item\label{step:nu-ini} Diagonalize $\tilde\sfH_\omega$ when $n_\nu^\tot$ is
large, and find the correspondence between $|\tilde\nu_l(\omega)\rangle$
and $|\nu_\alpha\rangle$ or $|\bar\nu_\alpha\rangle$
as was done in Table \ref{tab:nu-ini}.
\item\label{step:estimate-Es} Estimate 
the approximate locations of the swap points
as $n_\nu^\tot\rightarrow0$.
The swap points are expected to be in the neutrino sector
if the system is dominated by neutrinos instead of antineutrinos.
Pick a hierarchy for swap points, i.e., whether $\Ecatm>\Ecsol$
or $\Ecatm<\Ecsol$.
\item\label{step:nu-fin} Obtain from Eq.~\eqref{eq:omega123-0n} 
the correspondence between $|\tilde\nu_l(\omega)\rangle$
and $|\nu_i\rangle$ or $|\bar\nu_i\rangle$ for $n_\nu^\tot\rightarrow0$,
 as was done
in Table \ref{tab:nu-fin}.
\item\label{step:P} Use 
the results in step \ref{step:nu-ini} and \ref{step:nu-fin}
and the adiabatic ansatz in Eq.~\eqref{eq:equal-as} to find
neutrino conversion probabilities $P(\nu_\alpha\rightarrow\nu_i)$ and
$P(\bar\nu_\alpha\rightarrow\bar\nu_i)$, as was done in Table \ref{tab:P}.
\item Find lepton numbers $L_3$ and $L_8$ as $n_\nu^\tot\rightarrow0$. 
These will be functions of
 $\Ecsol$ and $\Ecatm$ when the initial energy spectra
$f_\nu(E)$ and the results in step \ref{step:P} are used.
\item\label{step:solve-Es} Solve 
for $\Ecsol$ and $\Ecatm$ by using lepton number conservation
$L_3|_{t=0}=L_3|_{n_\nu^\tot\rightarrow0}(\Ecsol,\Ecatm)$
and $L_8|_{t=0}=L_8|_{n_\nu^\tot\rightarrow0}(\Ecsol,\Ecatm)$.
If no consistent solution can be found, pick the other
hierarchy for swap points $\Ecsol$ and $\Ecatm$ in step \ref{step:estimate-Es}
and repeat steps \ref{step:nu-fin}--\ref{step:solve-Es}.
\end{enumerate}

\section{Adiabatic/Precession Solutions in a Dominant Matter
Background%
\label{sec:matt}}

\subsection{Effects of Neutrino-Electron Forward Scattering}

In the presence of ordinary matter, Eq.~\eqref{eq:eom} is
still valid except that the Hamiltonian for neutrino mode $\omega$
becomes
\begin{equation}
\sfH_\omega=\sfH_\omega^\vac
+\sfH^\matt+\mu\varrho_\tot,
\end{equation}
where $\sfH^\matt$ is the Hamiltonian contribution arising from
neutrino-electron forward scattering. In general this term
will give different
refractive indices for $\nu_e/\bar\nu_e$ and $\nu_{\mu,\tau}/\bar\nu_{\mu,\tau}$ 
\cite{Wolfenstein:1977ue,Wolfenstein:1979ni,Mikheyev:1985aa}.
Ignoring the trace term, in the flavor basis we can write
\begin{equation}
\sfH^\matt=\sqrt{2}\GF \nb\diag(Y_e,0,0),
\label{eq:Hmatt}
\end{equation}
where $\nb$ is the number density of baryons, and $Y_e$
is the electron fraction. MSW resonances
can occur if $n_e=\nb Y_e$ is small and comparable to $|\omega|/\sqrt{2}\GF$
or $|\kappa|/\sqrt{2}\GF$. In this section, however, we will assume
$n_e$ to be constant and very large for most neutrino modes:
\begin{equation}
n_e\gg\frac{|\kappa|}{\sqrt{2}\GF}.
\end{equation}

Because  $\sfH^\matt$ is independent of $\omega$, it vanishes 
in the corotating frame picked out by
\begin{equation}
\Hcr=\sfH^\matt.
\label{eq:cr-matt}
\end{equation}
In this corotating frame, the vacuum
Hamiltonian for neutrino mode $\omega$ becomes
\begin{widetext}
\begin{equation}
\tilde{\sfH}_\omega^\vac\simeq
\frac{\omega}{\Delta m_{21}^2}
\begin{pmatrix}
m_1^2c_{12}^2+m_2^2s_{12}^2
& \tilde{h}_{12}(t) &\tilde{h}_{13}(t)\\
\tilde{h}_{12}^*(t)
& \frac{1}{2}(m_3^2+m_1^2s_{12}^2+m_2^2c_{12}^2)
& \frac{1}{2}(m_3^2-m_1^2s_{12}^2-m_2^2c_{12}^2)\\
\tilde{h}_{13}^*(t)
& \frac{1}{2}(m_3^2-m_1^2s_{12}^2-m_2^2c_{12}^2)
& \frac{1}{2}(m_3^2+m_1^2s_{12}^2+m_2^2c_{12}^2)
\end{pmatrix}
\label{eq:HV-cr}
\end{equation}
\end{widetext}
in the flavor basis, where $c_{ij}=\cos\theta_{ij}$,
$s_{ij}=\sin\theta_{ij}$, and $\tilde{h}_{12}(t)$ and $\tilde{h}_{13}(t)$
are functions that oscillate with angular frequency $\sqrt{2}\GF n_e$.
In deriving Eq.~\eqref{eq:HV-cr}
we have taken $\theta_{13}\simeq0$ and $\theta_{23}\simeq\pi/4$.
Because 
\begin{equation}
\sqrt{2}\GF n_e\gg|\tilde{h}_{12}(t)|,|\tilde{h}_{13}(t)|,
\end{equation}
we expect $\tilde{h}_{12}(t)$ and $\tilde{h}_{13}(t)$ to average to 0
and, therefore, to have little effect on neutrino flavor evolution.

Setting $\tilde{h}_{12}(t)$ and $\tilde{h}_{13}(t)$ to 0, we can
diagonalize $\tilde\sfH_\omega^\vac$:
\begin{equation} 
\tilde\sfH_\omega^\vac\rightarrow
\frac{\omega}{\Delta m_{21}^2}\diag(m_1^{\prime2},m_2^{\prime2},m_3^{\prime2}),
\end{equation}
where
\begin{subequations}
\begin{align}
m_1^{\prime2}&\simeq c_{12}^2m_1^2+s_{12}^2m_2^2,\\
m_2^{\prime2}&\simeq s_{12}^2m_1^2+c_{12}^2m_2^2,\\
m_3^{\prime2}&\simeq m_3^2
\end{align}
\end{subequations}
are the effective mass-squared values for neutrino states
\begin{subequations}
\begin{align}
|\nu_1^\prime\rangle&\simeq|\nu_e\rangle,\\
|\nu_2^\prime\rangle&\simeq\frac{1}{\sqrt{2}}(|\nu_\mu\rangle-|\nu_\tau\rangle),
\\
|\nu_3^\prime\rangle&\simeq\frac{1}{\sqrt{2}}(|\nu_\mu\rangle+|\nu_\tau\rangle).
\end{align}
\end{subequations}

We note that the flavor density matrix $\tilde\varrho_\omega$ in the corotating
frame associated with $\Hcr=\sfH^\matt$  obeys an e.o.m.~similar to that 
obeyed by
$\varrho_\omega$ in  vacuum [Eq.~\eqref{eq:eom}], except for small
perturbations occurring 
on very short time scales ($\Delta t\sim\GF^{-1}n_e^{-1}$). 
The only difference is that
the presence of a large net electron background breaks
the ``degeneracy'' between
 $\nu_e/\bar\nu_e$ and $\nu_{\mu,\tau}/\bar\nu_{\mu,\tau}$.
Therefore, we can obtain adiabatic/precession solutions using the same
procedure listed in Sec.~\ref{sec:adiabatic} but with the replacements
\begin{equation}
|\nu_i\rangle\longrightarrow|\nu_i^\prime\rangle
\quad\text{and}\quad 
m_i^2\longrightarrow m_i^{\prime2}.
\end{equation}
Likewise, the conserved lepton numbers should be calculated in the
$|\nu_i^\prime\rangle$ basis instead of the $|\nu_i\rangle$ basis.

\subsection{Effects of Virtual Charged Leptons}

At very large matter density, virtual $\mu$ and $\tau$ states
contribute to a higher order correction to neutrino refractive indices,
and $\sfH^\matt$ is found to be 
\cite{Fuller:1987aa,Botella:1987aa,Roulet:1995qb}
\begin{equation}
\sfH^\matt=\sqrt{2}\GF\nb\diag(Y_e,0,Y_\tau)
\end{equation}
in the flavor basis,
where $\nb Y_\tau$ gives the effective net $\tau$ lepton abundance. 
If the matter
density is so large that 
\begin{equation}
\sqrt{2}\GF \nb Y_\tau\gg|\kappa|,
\end{equation} 
we can again employ the corotating frame as 
in Eq.~\eqref{eq:cr-matt}. Ignoring  the rapidly-oscillating off-diagonal
elements diagonalizes $\tilde{\sfH}_\omega$  in the flavor basis
in this corotating frame:
\begin{equation}
\tilde{\sfH}_\omega^\vac\simeq
\frac{\omega}{\Delta m_{21}^2}
\diag(m_1^{\prime\prime2},m_2^{\prime\prime2},m_3^{\prime\prime2}),
\end{equation}
where
\begin{subequations}
\label{eq:m123-matt}
\begin{align}
m_1^{\prime\prime2}&\simeq c_{12}^2m_1^2+s_{12}^2m_2^2,\\
m_2^{\prime\prime2}&\simeq c_{23}^2(s_{12}^2m_1^2+c_{12}^2m_2^2)+s_{23}^2m_3^2,\\
m_3^{\prime\prime2}&\simeq s_{23}^2(s_{12}^2m_1^2+c_{12}^2m_2^2)+c_{23}^2m_3^2
\end{align}
\end{subequations}
are the effective mass-squared value for the neutrino states
\begin{subequations}
\begin{align}
|\nu_1^{\prime\prime}\rangle&\simeq|\nu_e\rangle,\\
|\nu_2^{\prime\prime}\rangle&\simeq|\nu_\mu\rangle,\\
|\nu_3^{\prime\prime}\rangle&\simeq|\nu_\tau\rangle.
\end{align}
\end{subequations}
Therefore, the adiabatic/precession solution still obtains in the presence
of both large electron and large effective tau abundances.
In this case, however, the degeneracy among 
$\nu_e/\bar\nu_e$, $\nu_\mu/\bar\nu_\mu$
and $\nu_\tau/\bar\nu_\tau$ is completely broken.
In this case, the adiabatic/precession solution is best obtained
in the flavor basis. 
The conserved lepton numbers should also be calculated in the flavor basis.

Because the absolute
neutrino masses are irrelevant for neutrino oscillations,
in  the inverted neutrino mass hierarchy case
we can take $m_1\simeq m_2^2\simeq \dmatm>m_3^2=0$.  
When $\nb Y_\tau$ is large,
from Eq.~\eqref{eq:m123-matt}
it can be seen that the mass-squared eigenvalue
for $|\nu_1^{\prime\prime}\rangle$
is always heavier than the mass-squared eigenvalues
for $|\nu_2^{\prime\prime}\rangle$ and 
$|\nu_3^{\prime\prime}\rangle$,
while the effective $23$-mass-hierarchy, or the sign of
\begin{equation}
\Delta m_{32}^{\prime\prime2}\equiv m_3^{\prime\prime2}-m_2^{\prime\prime2},
\end{equation}
depends on whether $\theta_{23}$ is larger or smaller than $\pi/4$.
Similarly, for the normal neutrino mass hierarchy case, we can
take $m_3^2\simeq\dmatm>m_1^2\simeq m_2^2\simeq0$, and again
the effective $23$-mass-hierarchy depends on 
 whether $\theta_{23}$ is larger or smaller than $\pi/4$,
although in a reversed fashion. According to the discussion
in Sec.~\ref{sec:swapping}, the final energy spectra for
$\nu_2^{\prime\prime}/\bar\nu_2^{\prime\prime}$ and
$\nu_3^{\prime\prime}/\bar\nu_3^{\prime\prime}$ 
(and, therefore, the spectra for $\nu_\mu/\bar\nu_\mu$ and
$\nu_\tau/\bar\nu_\tau$) 
interchange with each other when $\theta_{23}$ rotates
from the first octant to the second octant. In other
words, the final neutrino energy spectra can be
sensitive to deviations from maximal 23-mixing.
This can be illustrated using the toy model discussed in 
Ref.~\cite{EstebanPretel:2007yq}.

In this toy model the neutron star emits only $\nu_e$ and
 $\bar\nu_e$ with the same energy $E_\nu$ into a
thick matter envelope where $\nb Y_\tau$ is large.
In this model it is  assumed  also that $n_{\nu_e}/n_{\bar\nu_e}=1+\epsilon$
at the neutrino sphere with $\epsilon>0$.
If the neutrino gas follows the adiabatic/precession
solution, then $P(\nu_e\rightarrow\nu_i^{\prime\prime})$
and $P(\bar\nu_e\rightarrow\bar\nu_i^{\prime\prime})$ must be either
0 or 1 except at swapping points (see Sec.~\ref{sec:swapping}).
However, it is clear from the conservation of lepton numbers
in the flavor basis that $\nu_e$'s can not have been fully converted
to other flavors. Because neutrino (instead of antineutrino)
is the dominant species in the system, the spectral swaps
occur in the neutrino sector.
Therefore, $\bar\nu_e$'s are completely
converted into antineutrinos of another flavor
as $n_\nu^\tot\rightarrow0$. 
In this case, the precession ansatz
in Eq.~\eqref{eq:prec-ansatz} is trivially satisfied 
as $n_\nu^\tot\rightarrow0$
in the corotating frame defined by
\begin{equation}
\Hcr=\sfH^\vac_{\omega_+},
\end{equation}
where $\sfH^\vac_{\omega_+}$ is the vacuum term in the Hamiltonian
for the neutrino mode with energy $E_\nu$. In other words,
both swapping points have collapsed
into one which is located in the neutrino sector and at energy
$E^\mathrm{s}=E_\nu$. Following the discussions in Sec.~\ref{sec:swapping},
we can use the conservation of
lepton numbers in the flavor basis to find that
 the neutrino conversion
probabilities are
\begin{subequations}
\label{eq:P-matt}
\begin{align}
P(\nu_e\rightarrow\nu_e)&=\frac{\epsilon}{1+\epsilon},\\
P(\nu_e\rightarrow\nu_\mu)&=0,\\
P(\nu_e\rightarrow\nu_\tau)&=\frac{1}{1+\epsilon},\\
P(\bar\nu_e\rightarrow\bar\nu_e)&=0,\\
P(\bar\nu_e\rightarrow\bar\nu_\mu)&=0,\\
P(\bar\nu_e\rightarrow\bar\nu_\tau)&=1
\end{align}
\end{subequations}
for the inverted neutrino mass hierarchy and $\theta_{23}<\pi/4$.
For the inverted neutrino mass hierarchy and $\theta_{23}>\pi/4$
we can obtain  neutrino conversion probabilities which are similar to
those in Eq.~\eqref{eq:P-matt} but with $\nu_\mu\leftrightarrow\nu_\tau$
and $\bar\nu_\mu\leftrightarrow\bar\nu_\tau$.
 This is exactly what has been observed in the numerical calculations
for the toy model at $r\simeq400$ km 
(Fig.~2 in Ref.~\cite{EstebanPretel:2007yq}).

\section{Conclusions%
\label{sec:conclusions}}

We have developed a framework for studying collective
three-flavor neutrino oscillations. Important techniques in
studying collective two-flavor oscillations such as corotating frames
can be applied readily  to three-flavor scenarios in 
this framework. We have shown that the three-flavor
adiabatic/precession solution obtains when both the precession ansatz and
the adiabatic ansatz are satisfied. If the flavor
evolution of a neutrino gas is described by the adiabatic/precession solution,
the final neutrino energy spectra will exhibit the stepwise swapping
phenomenon. We have shown that stepwise spectral swapping
appears in a numerical example in which neutrinos are directly
emitted from the neutrino sphere of a bare neutron star into vacuum.
For this special example, because the neutrinos in $\mu$ and $\tau$ flavors
are equally mixed and have identical energy spectra initially,
the adiabatic/precession solutions for both the three-flavor and the two-flavor
scenarios produce the same spectral swapping at the
atmospheric neutrino mass-squared-difference scale.
 In more general cases, however, this
may not be the case, and the full $3\times3$
mixing framework should be employed.

Strictly speaking, the adiabatic/precession solution
obtains  only when the $CP$-violating phase $\delta=0$. This is because,
if $\delta\neq0$, the unitary transformation matrix $U$ connecting the
flavor states and vacuum mass eigenstates is not real, and
\begin{equation}
U\rho_\bfp^*U^\dagger\neq(U\rho_\bfp U^\dagger)^*
\text{ and }
U\bar\rho_\bfp^*U^\dagger\neq(U\bar\rho_\bfp U^\dagger)^*.
\end{equation}
As a result, Eqs.~\eqref{eq:eom-nu} and \eqref{eq:eom-anu} become
invalid in the vacuum mass basis, and all the following derivations are invalid.
In practice, however, the adiabatic/precession solution may still be a good
approximation even when $\delta$ is large.
This is because $\theta_{13}\simeq0$ and the transformation
matrix $U$ is almost real.  In fact,
numerical simulations in Ref.~\cite{Duan:2007sh} show that,
at least for the parameters employed in those simulations,
varying the $CP$ phase $\delta$ has little effect on the final
neutrino energy spectra except for changing the relative mixing
of $\mu$ and $\tau$ neutrino flavors.
The possible effects of $CP$ violation in stellar collapse
have been discussed in a different context
(see, e.g., Ref.~\cite{Balantekin:2007es}).

We also have  demonstrated that the adiabatic/precession solution obtains
even in the presence of a dominant matter background and a large
mu-tau term. When the matter term is much larger than the vacuum
mixing term, the presence of the ordinary matter only reshuffles
the neutrino states in which the spectral swapping  has the
most dramatic manifestation. For the supernova environment, this means
that the regime where collective neutrino oscillations occur
is solely determined by the neutrino fluxes and does not 
depend sensitively on the matter density profile.
This is in agreement with the previous analysis in
 two-flavor scenarios \cite{Duan:2005cp}. The supernova neutrino
signals observed on earth will depend of course  on the matter
profile in the supernova.
In part, this is because the MSW effect will
modify the neutrino energy spectra subsequent to
the collective oscillations discussed here.
This dependence of neutrino signals on the matter profile
can provide important information on the 
conditions deep in the supernova envelope 
\cite{Schirato:2002tg,Duan:2007sh,Lunardini:2007vn}.

\begin{acknowledgments}
This work was supported in part by 
DOE grants DE-FG02-00ER41132 at INT,
DE-FG02-87ER40328 at UMN,
NSF grant PHY-04-00359 at UCSD,
and an IGPP/LANL mini-grant.
\end{acknowledgments}

\bibliography{ref}

\end{document}